Estimation of double differential angle-dependent neutron production cross sections from tritons on $^{197}$Au at energies from 5.97 to 19.14 MeV

Manfred Drosg

Fakultät für Physik der Universität Wien

Institut für Experimentalphysik

Strudlhofgasse 4

A-1090 Wien

Austria

phone:+43 1 427751001

email Manfred.Drosg@univie.ac.at

Bernard Hoop (ret.)

Pulmonary and Critical Care Unit

Department of Medicine

Massachusetts General Hospital

Harvard Medical School

Boston, MA 02114

USA

phone: 207 439 1636

email: bhoop@earthlink.net

**Abstract.** Estimated cross sections for neutron production from triton bombardment of gold are deduced from measurements of triton interactions with gas targets that used gold as a triton beam stop material. Differential cross sections for production of neutrons from 5.97-, 7.47-, 10.45-, 16.41- and 19.14-MeV tritons on $^{197}$Au were evaluated. Corrections for the neutron interaction in gold, in the target structure and in the air of the flight path were obtained by means of a Monte Carlo technique. Uncorrelated scale uncertainties range from 24 to 41% whereas those of double differential cross sections range from 0.2 to 5%. Based on these cross section data, calculation of neutron yield at 0º from fully stopped tritons at 20.22 MeV agree with an independent measurement. Least-squares fits with a gamma distribution model indicate an anisotropy in the high energy portion of the neutron spectra. Legendre polynomial fits of differential cross sections are reported. All neutron cross section data are made available through the Experimental Nuclear Reaction Data (EXFOR) library at international data centers.

# I. INTRODUCTION

Neutrons produced in interactions of tritons with the three hydrogen isotopes and with $^4$He are a characteristic feature in primordial and stellar nucleosynthesis, in applications of nuclear fusion, and may be of interest in *ab initio* theory of light ion reactions [1]. Measurements of these cross sections are usually made using gaseous targets. A principal difficulty encountered in making such measurements is correcting for background neutrons produced by tritons that interact with the target structure [2].

For gas targets, background measurements may be performed with 1) a target cell filled with an inert gas of equivalent projectile stopping power, or 2) an empty target cell. In the latter case, neutron background from triton interactions with entrance foil material can be properly corrected, but correction for neutron background from triton interactions with beam stop material is overestimated, due to the missing attenuation of the triton energy in the gas target. Thus, the difference between neutron production from an empty cell and from that of a cell filled with an inert dummy gas can be deduced from neutron production from tritons on gold with an effective gold target thickness given by the energy loss in the target gas. Hence, double-differential neutron emission cross sections of gold can be extracted.

For example, measurements of $^3$H(t,n) double-differential cross sections were made using a gas cell with a molybdenum foil as entrance window and a gold disc as beam stop. In this example, an inert gas target filling – hydrogen – was available for correction of the neutron background at "back angles." That is, $^1$H(t,n) produces neutrons into a forward cone only, a process called "kinematic collimation". Thus, for triton energies below breakup threshold of 25.01 MeV, no neutrons are produced heading outside of this cone. This kinematic collimation of neutron production occurs because the target mass is smaller than the projectile mass. Hence, the velocity of the emitted neutron is smaller than that of the center-of-mass (c.m.) velocity, such that neutron emission at all c.m. angles occurs inside a forward cone with a half opening angle of less than 90 degrees in the laboratory system, i.e., no neutrons are produced heading outside of this cone.

This half-opening angle increases with energy, but remains, for tritons on $^1$H, smaller than 90 degrees at all energies.

In the course of measuring neutron production cross sections of tritons on $^2$H, $^3$H, and $^4$He, background neutron spectra were obtained either by measuring, under identical conditions, neutron output from empty target cell, or at larger angles (≥60°) with hydrogen in the target cell.

By measuring at the same angle, the shapes of both types of background spectra may be compared. A comparison is also possible over those portions of the monoenergetic $^1$H(t,n)$^3$He and $^2$H(t,n)$^4$He spectra which are devoid of intrinsic neutrons, i.e. beyond the energy of neutrons from two-body reaction, and in the case of t-$^2$H, in the energy interval between two-body peak and neutron distribution from break-up reactions.

From a number of such comparisons, it became clear that the yield of the empty-cell background spectra and those from cells filled with either hydrogen or deuterium is greater than the statistical uncertainty. The expected difference at the high energy end of the neutron spectra could not be seen due to insufficient intensity resulting in poor statistical quality of these higher energy neutron data. Thus, the excess neutrons in the empty-cell case reflects neutron production by a gold target of the same thickness (expressed in energy difference) as the gas target, and at the same nominal energy. This unusual approach deserves further explanation. First of all, the energy deposited by the triton beam in the gas is very small (1.4% at 5.97 MeV and just 0.1% at 19.14 MeV). Thus, it is not surprising that there is no dramatic change in the shape of the background spectrum in the gas-in and the gas-out case. By substituting the gas filling by a gold foil of equivalent thickness (between 1.79 and 0.81 mg/cm$^2$) the tritons impinging on the beamstop would have (to a first order) the same characteristics as those having travelled through the gas. As the tritons are fully stopped in the beamstop the first layer of it fulfils perfectly the task of such a gold foil. Actually, the neutron yield comparison is not made with a target filled with an inert gas (for the background determination) vs. an empty target but vs. a target with the gas replaced by a gold foil of appropriate thickness. Thus, the measured quantity is really the yield from the gold foil (even if imaginary, i.e. not present as such) that can be converted into cross sections.

Although it is not possible to reliably quantify the accuracy of this subtraction procedure, we believe the uniqueness of these cross section results constitutes a useful contribution to nuclear data even if we call these data estimated data.

## II. EXPERIMENTS

Experiments were performed at the HVEC tandem accelerator of what was formerly known as the Ion Beam Facility (IBF) at Los Alamos National Laboratory, using a time-of-flight system with a 2" x 2" liquid scintillator (NE213) as a neutron detector. Two pulse-height biases (minimum between the 25 keV and the 60 keV peak in the $^{241}$Am gamma spectrum, and Compton edge of $^{137}$Cs) were used simultaneously; the flight path was 3.61 m, the total time resolution was 2.1 ns (FWHM of the prompt gamma-ray peak). The heavily shielded detector could be rotated around a pivot situated under the center of the gas target. The angles were measured with a resolution of 0.1 degree. The strong angle dependence of the $^1$H(t,n)$^3$He cross section allowed an experimental determination of the actual 0 degree position to be $(0.30 \pm 0.05)$ degree by measuring at the same nominal angle left and right.

The energy dependence of the efficiency of the neutron detector at the two biases was determined by 1) measuring neutron yields of the reactions $^2$H(t,n)$^4$He and $^1$H(t,n)$^3$He at about 30 neutron energies between 2.2 and 36.9 MeV, and 2) comparing them with the well-known cross sections [3]. Gas targets of 3.0 cm length with an entrance foil of 5.3 mg/cm$^2$ molybdenum and a beam-stop made of gold (0.076 cm thick) were used [4,5]. The digital recorder of the gas pressure (about 0.2 MPa) made it possible to have (within 0.4%) the same gas pressure independent of the filling. By comparing the measured yield from the $^1$H(t,n)$^3$He reaction with the calculated yield [3], the experimental setup was calibrated to better than 5%. At five energies, namely 5.97, 7.47, 10.45, 16.41, and 19.14 MeV, measurements with hydrogen filling and with empty cells were available at the same angles, so that at these five energies, double-differential neutron production cross sections for tritons on $^{197}$Au could be extracted.

## III. DATA ANALYSIS

The data essential to this evaluation are the ratio of neutrons measured with an empty gas target to those measured (under otherwise identical conditions) when the target is filled with an inert gas (hydrogen). The quality of the extracted data depends on the following two assumptions: 1) changes in the shape of the two types of neutron spectra can be disregarded, and 2) the ratio of all neutrons produced in a gas-filled target to that of an empty target is independent of the angle. It is evident that the first assumption is only a crude approximation albeit the small energy loss in the gas (<1.4%). For kinematic reason alone, a pronounced difference in the high energy portion would be expected. However, its contribution is less than the statistical uncertainty.

The second assumption is valid as long as there is no significant change in the shape of the angular yield distribution with a change in energy corresponding to the energy loss in the target (<0.084 MeV). From the data analysis it became clear that the gas-in spectra contained, as could be expected, more neutrons from the room background than the gas-out spectra. This background stems mainly from primary source neutrons inscattered from the air volume seen by the detector. As the reaction $^1$H(t,n)$^3$He has, in this energy range, typically a seven times higher primary yield than $^2$H(t,n)$^4$He this effect is more serious for t-H than for t-D. Besides, as the absolute value of this type of background does not change much with triton energy, its effect is strongest at the lower triton energies where the gold cross sections are smaller.

Table I shows the raw data: the yield ratio between the spectrum of the empty cell and that of a cell filled with hydrogen, under otherwise identical conditions. The effective gold target thickness is the depth in gold at which the triton beam has lost the same amount of energy as in the gas of the gas-filled cell. Differences in angular and energy straggling between equivalent layers of hydrogen and gold are expected not to affect the results by any noticeable amount, The values called "net" give the fraction of the gas-out spectrum that stems from the first layer in gold which is as thick (measured in energy loss) as the hydrogen filling. The column *scale uncert.* gives the uncorrelated part of the scale uncertainty.

Table I. Raw data

| $E_t$ | Ratio | Δratio | net | Δnet | $\Delta E_{Au}$ | $\Delta E_H$ | $f_Y$ | Net*$f_Y$ | scale uncert. |
|---|---|---|---|---|---|---|---|---|---|
| 5.97 | 1.293 | 0.154 | 0.227 | 0.092 | 45.73 | 390.1 | 22.91 | 5.19 | 41% |
| 7.47 | 1.220 | 0.103 | 0.180 | 0.069 | 40.79 | 324.9 | 24.54 | 4.43 | 38% |
| 10.45 | 1.115 | 0.034 | 0.104 | 0.026 | 34.06 | 246.5 | 27.00 | 2.79 | 25% |
| 16.41 | 1.0171 | 0.0036 | 0.0187 | 0.0045 | 26.22 | 169.2 | 30.29 | 0.566 | 24% |
| 19.14 | 1.0197 | 0.0055 | 0.0212 | 0.0063 | 23.89 | 148.8 | 31.38 | 0.665 | 30% |

For conversion of neutron yields from gold, which had been normalized by means of the $^1H(t,n)^3He$ cross section, to differential cross sections, the factor $f_Y = (m_{Au} \cdot \Delta E_{Au})/(m_H \cdot \Delta E_H)$ was used, where $m_{Au}$ and $m_H$ are molar masses, and $\Delta E$ are specific energy losses $dE/dx$ taken from PSTAR [6]. The source of the neutron spectra is inside the beam-stop, a cylindrical disc of gold perpendicular to the beam, 0.076 cm thick with a diameter of 1.0 cm. In particular, at 90 degrees self-attenuation is an issue. In addition, neutron interactions with air in the flight-path of 3.6 m must be taken into consideration. Consequently, the necessary corrections were determined by means of a Monte Carlo simulation using MCNPX [7]. This simulation covered neutron interactions with the target structure and the air in the fight path. Maximum correction occurred, as expected, at 90 degree. There, the correction amounted to about 12 %.

Table II. Angle-dependent differential neutron production cross sections with their uncertainties in mb/sr for neutrons with energy > 0.5 MeV, unless otherwise indicated. Tabulated uncertainties (Unc.) are statistical only.

| $E_t$(MeV) angle\ (deg) | 5.97 | Unc. | 7.47 | Unc. | 10.45 | Unc. | 16.41 | Unc. | 19.14 | Unc. |
|---|---|---|---|---|---|---|---|---|---|---|

| | | | | | | | | | |
|---|---|---|---|---|---|---|---|---|---|
| 0 | 27.60 | 0.34 | 57.51 | 0.47 | 109.6 | 0.5 | 464.6 | 1.2 | 987.53 | 0.94 |
| 0 [1] | 13.78 | 0.23 | 27.13 | 0.36 | 56.91 | 0.38 | 241.54 | 0.87 | 517.01 | 0.73 |
| 15 | 25.61 | 0.31 | 54.80 | 0.51 | 104.9 | 0.5 | 442.68 | 0.60 | - | - |
| 30 | 23.72 | 0.34 | 50.95 | 0.43 | 99.31 | 0.45 | 401.62 | 0.56 | 859.839 | 0.86 [2] |
| 50 | - | - | - | - | - | - | 373.90 | 0.70 | - | - |
| 60 | 25.40 | 0.47 | 49.84 | 0.41 | 92.95 | 0.45 | 353.74 | 0.70 | 741.76 | 0.79 |
| 90 | 23.72 | 0.47 | 48.09 | 0.43 | 87.00 | 0.45 | 300.59 | 0.49 | 661.76 | 0.78 |
| 120 | 19.47 | 0.41 | 39.93 | 0.38 | 76.82 | 0.40 | 299.43 | 0.47 | 650.65 | 0.74 |
| 145 | 18.33 | 0.39 | 35.91 | 0.36 | 48.90 | 0.28 | 288.84 | 0.46 | 637.60 | 0.73 |
| $\sigma$ [3] | *0.27* | *0.11* | *0.57* | *0.22* | *1.02* | *0.25* | *4.10* | *0.99* | *8.84* | *2.63* |

[1] neutrons with energy > 2 MeV only

[2] This small statistical uncertainty just means that $10^6$ counts have been recorded in this spectrum

[3] cross sections integrated over angles, in barn

A total of 35 double-differential neutron emission cross section spectra were extracted. Integration over neutron energy yields angle-dependent differential cross sections. Table II summarizes these and angle integrated cross sections at the five triton energies. These data are the measured cross sections for the emission of neutrons with energy greater than 0.5 MeV. The second line at 0 degrees gives the cross sections of emitting neutrons with an energy > 2.0 MeV, discussed in Section IV. The last line gives the angle-integrated cross sections. These are displayed as solid circles on the left in Fig. 1. Exponential energy dependence is due to the fact that with rising energy the multiplicity of emitted neutrons increases markedly, as indicated for the reactions listed in Table III. The threshold energies beyond 19.14 MeV show that several more exit channels containing neutrons are opened when the triton energy is increased to about 24 MeV.

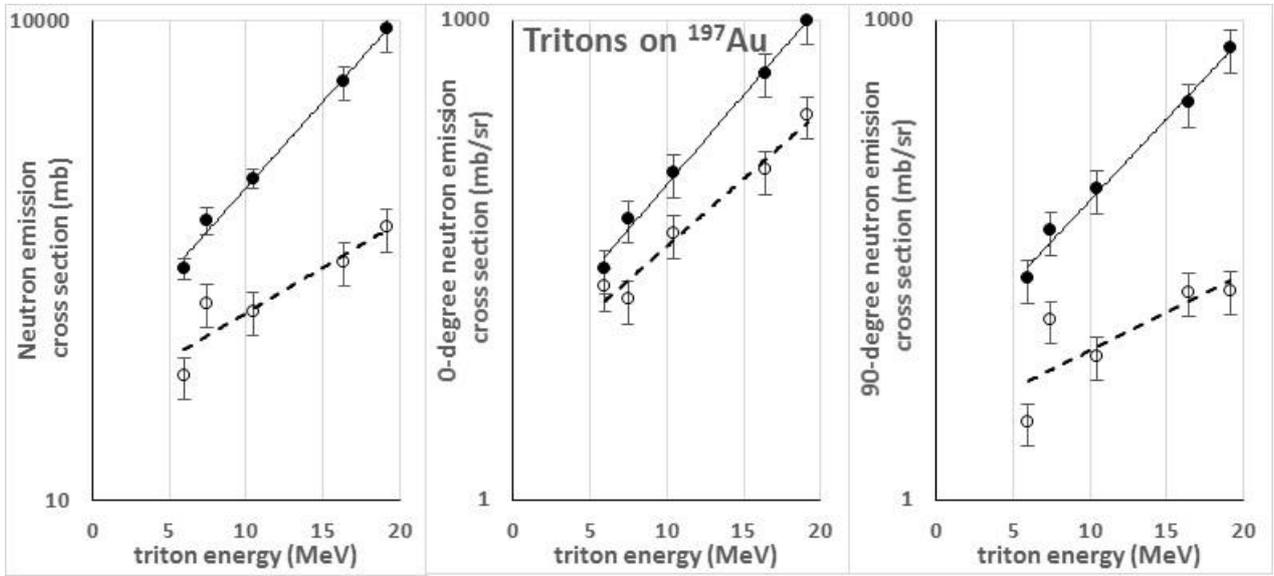

Fig. 1. Energy dependence of neutron emission cross section for tritons on $^{197}$Au for neutrons with energy of greater than 0.5 MeV. Measured data are shown by solid circles. Cross section for high energy component is shown by open circles. Exponential fits to the data are shown by solid lines.

Table III. Some examples of neutron multiplicity together with negative $Q$-values, unless shown otherwise. Quantities are in MeV. Threshold energies of reactions are 1.531% higher than (negative) $Q$-values. Values are obtained via a $Q$-value calculator [8].

|  | +n | +2n | +3n | +4n | +5n | +6n |
|---|---|---|---|---|---|---|
| t, | +5.2719 | 1.3920 | 10.8707 | 16.6624 |  |  |
| t,p | 1.9695 | 8.4818 | 16.5542 | 23.1959 |  |  |
| t,d | 6.2572 | 14.3296 | 20.9714 |  |  |  |
| t,t | 8.0724 | 14.7141 | 23.1418 |  |  |  |
| t,$^3$He | 6.5341 | 14.4687 | 20.5738 |  |  |  |
| t,$^4$He | +6.1216 | +0.1656 | 8.3409 | 14.5963 | 23.2712 |  |
| t,$^6$Li | 3.8111 | 11.5830 | 17.7812 |  |  |  |
| t,$^9$Be | +3.0985 | 4.6938 | 10.6143 | 18.6039 |  |  |
| t,$^{12}$C | +15.6438 | +10.1770 | +2.9848 | 2.7689 | 10.1805 | 16.3714 |

## IV. DISCUSSION

As these data estimates are unique, no direct comparison with other data is possible. However, there exists an independently measured neutron yield at 0 degrees for fully stopped tritons in gold [9] with a neutron energy bias of 2.0 MeV. Using the present zero degree differential cross section data for neutron energies above 2.0 MeV (Table II) at the five triton energies and interpolating semi-logarithmically between the triton energies, the yield for fully stopped triton beams up to 19.14 MeV can be calculated using the energy loss values of PSTAR [6]. Extrapolation from 19.14 MeV to 20.22 MeV used the same slope as for the interpolation between 16.41 MeV and 19.14 MeV. Thus, a 0-degree yield of neutrons with energy > 2 MeV for fully stopped 20.22 MeV tritons in gold of 653 n/(sr*pC) was obtained with an uncertainty in excess of 144 n/(sr*pC). The close agreement of this value with measured value [9] of (616 ± 31) n/(sr*pC) supports the chosen procedure. Although threshold for neutron detection in the experiment is ca. 0.29 MeV, the slope of neutron detection efficiency curve prohibits extraction of reliable data below a neutron energy of about 0.5 MeV. Thus a neutron evaluation threshold energy of 0.5 MeV was used.

Resulting double differential cross sections in units of mb/(sr*MeV) at 0° and at 90° are plotted in Fig. 2.

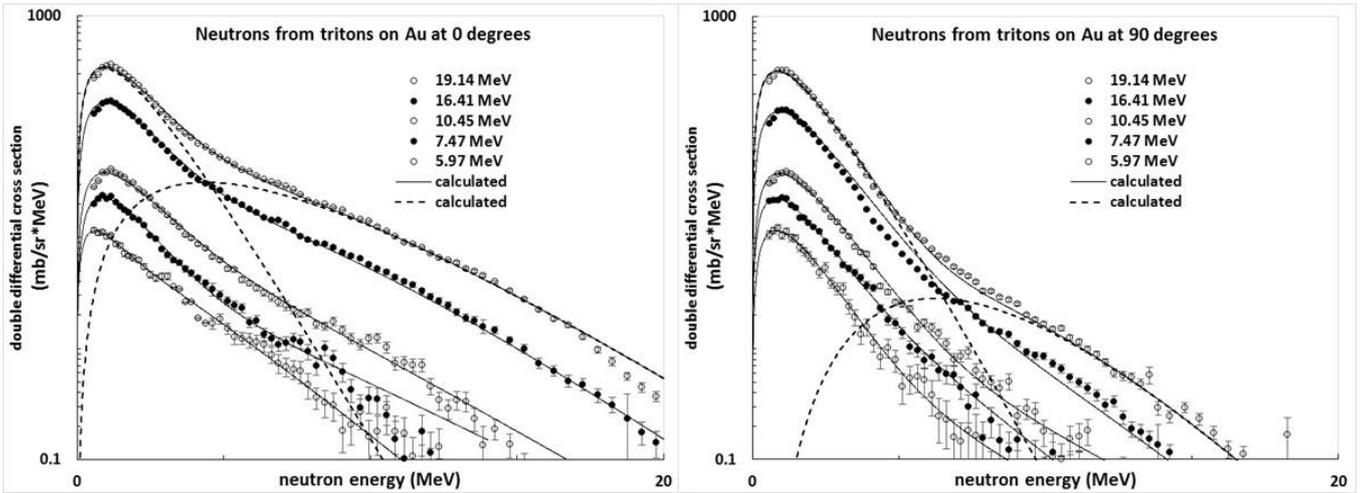

Figure 2. Estimated double differential neutron production cross sections for neutrons at 0 and 90 degree from tritons of 5.97 to 19.41 MeV on $^{197}$Au. Calculated cross sections from gamma distribution fits are shown by solid curves. Two gamma distribution components that make up the 19.14-MeV fit are shown by dashed curves. Error bars include statistical uncertainties, only.

In Fig. 2, measured spectra expressed as double differential cross sections $d^2\sigma/dEd\Omega$ are represented by a sum of two gamma distributions expressed in the general form, $N \cdot E^{k-1}\theta^{-k}[\Gamma(k)]^{-1}e^{-E/\theta}$, where $N$ is a normalization constant, $E$ is neutron energy, $\theta$ and $k$ are, respectively, scale and shape parameters, and $\Gamma(k)$ is the Gamma function for value $k$. The low energy portion of neutron spectra obtained at all triton energies is found to be adequately described at each angle by a distribution for which $k = 2$. The high energy portion of neutron spectra obtained at 5.97, 7.47, and 10.45 MeV triton energies are adequately described at each angle by a distribution for which $k = 1.5$ [10,11].. A relatively small portion of neutron spectra at all angles at 16.41 and 19.14 MeV is described by gamma distributions with $3 \leq k \leq 6$. By way of illustration, the two gamma distribution components of the 19.14-MeV fit are shown by dashed curves in Fig. 2.

Parameters $N_i$ and $\theta_i$ for $i = 0$ and 1 components of gamma distribution fits to double differential neutron production cross section are given in Table IV. Corresponding values of $k$ are also indicated in Table IV.

Table IV. Parameters $N_i$ (in mb/(sr*MeV)) and $\theta_i$ (in MeV) for $i = 0$ and 1 components of gamma distribution fits to double differential neutron production cross sections for neutrons with energy > 0.5 MeV (cf. IV. Discussion).

| $E_t$ (MeV)\ Angle (deg) | 5.97 | | 7.47 | | 10.45 | | 16.41 | | 19.14 | |
|---|---|---|---|---|---|---|---|---|---|---|
| | 0 [1] | 1 [2] | 0 [1] | 1 [2] | 0 [1] | 1 [2] | 0 [3] | 1 [2] | 0 | 1 [2] |
| 0 | 23.8 | 8.1 | 19.5 | 45.2 | 49.2 | 72.9 | 116.8 | 391.2 | 256.3 | 830.4 |
| | 1.9 | 0.51 | 3.0 | 0.83 | 2.67 | 0.90 | 2.00 | 0.90 | 2.20 [3] | 0.90 |
| 15 | 17.2 | 12.3 | 23.5 | 39.2 | 33.9 | 81.8 | 90.4 | 400.0 | - | - |
| | 2.0 | 0.66 | 2.56 | 0.75 | 2.8 | 0.95 | 2.10 | 0.90 | | |
| 30 | 11.9 | 16.0 | 15.2 | 42.6 | 21.8 | 87.5 | 63.8 | 380.3 | 135.5 | 814.9 |
| | 2.3 | 0.64 | 2.92 | 0.80 | 2.80 | 0.95 | 1.95 | 0.90 | 1.63 [4] | 0.90 |
| 50 | - | - | - | - | - | - | 46.0 | 371.7 | - | - |
| | | | | | | | 1.95 | 0.90 | | |
| 60 | 3.4 | 25.5 | 18.2 | 37.9 | 17.4 | 86.5 | 32.2 | 363.3 | 48.7 | 785.0 |

|     | 2.6 | 0.88 | 2.00 | 0.82 | 2.80 | 0.90 | 1.90 | 0.90 | 1.33 [5)] | 0.90 |
| --- | --- | --- | --- | --- | --- | --- | --- | --- | --- | --- |
| 90  | 3.4 | 24.0 | 14.7 | 40.5 | 8.4 | 90.0 | 19.9 | 323.0 | 20.3 | 733.4 |
|     | 2.6 | 0.80 | 1.93 | 0.78 | 2.80 | 0.90 | 1.70 | 0.86 | 1.25 [5)] | 0.84 |
| 120 | 3.4 | 19.0 | 13.4 | 32.7 | 9.7 | 77.8 | 8.6 | 333.0 | 8.5 | 735.3 |
|     | 2.6 | 0.80 | 1.74 | 0.76 | 2.7 | 0.87 | 1.90 | 0.86 | 1.30 [5)] | 0.84 |
| 145 | 3.4 | 17.5 | 11.4 | 29.7 | 8.4 | 47.3 | 3.4 | 325.6 | 10.0 | 720.2 |
|     | 2.6 | 0.80 | 1.73 | 0.75 | 2.6 | 0.85 | 2.60 | 0.85 | 1.40 [5)] | 0.83 |

[1)] $k = 1.5$; [2)] $k = 2.0$; [3)] $k = 3.0$; [4)] $k = 4.0$; [5)] $k = 6.0$

Measured spectra were fitted by means of an optimized version of the Levenberg-Marquardt method for minimization by default (*genfit* under PTC Mathcad 3.0), and total cross section at each angle is calculated as the integral of the best-fit function over the entire energy interval. Calculated spectra at 0º and 90º over neutron range 0 to 20 MeV are shown by solid curves in Fig. 2.

Calculated cross sections are compared with Legendre polynomial fits in Figure 3.

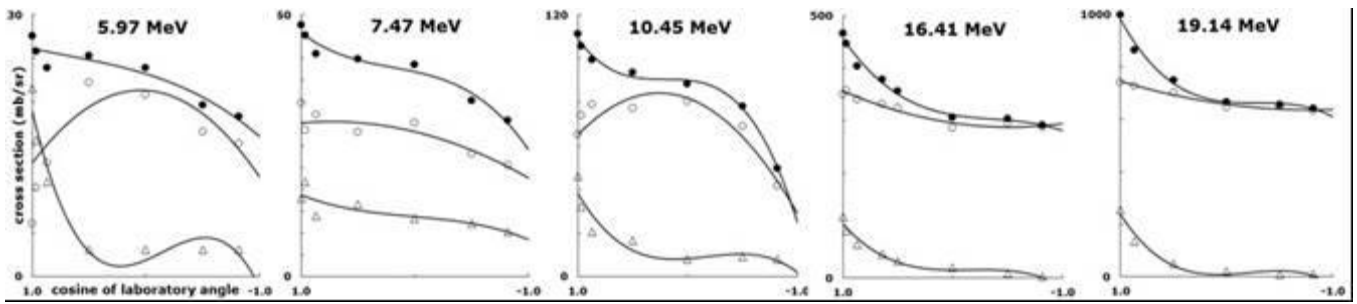

Figure 3. Calculated differential cross section at each angle (solid circles) for the five energies of the experiment. Measured cross section uncertainties (Table II) lie within the size of the plotted solid circles. Cross sections for the two gamma distribution components are also shown (open circles and triangles, respectively). Fourth-order Legendre polynomials are shown as solid curves.

Significant anisotropy in angular distributions of both $k = 1.5$ and $k = 2$ components is present in the three lower energy cross sections (Fig. 3, left three panels). In 16.41- and 19.14-MeV cross sections, forward-peaked anisotropy is evident in the high energy portion of the neutron spectra (Fig. 3, right two panels). It should be noted that, of the two gamma distribution components (dashed curves in Fig. 2) that comprise the fit to the 0- and 90-degree 19.14-MeV data (solid curves in Fig. 2), the lower energy component is the largest contribution to the calculated cross section.

In a study of neutron evaporation spectra from proton bombardment of Au at energies between 6 and 12 MeV, continuous spectra resulted which had a Maxwellian energy distribution. The nuclear temperature showed, however, an anomalous increase with bombarding energy [12]. Although the fitted functions in the present work adequately describe neutron spectra over nearly four decades of the cross section scale at all angles, systematic discrepancy between calculated and measured distribution (cf. Fig. 2) may very well be the inadequacy of the chosen sum of gamma distributions to represent the spectra. Limitations in statistical quality of the data, especially at high neutron energies, preclude a more comprehensive description.

Although present cross sections are adequately fitted at all angles and triton energies by a sum of two gamma distributions involving four parameters (i.e., $N_0$, $\theta_0$, $N_1$, $\theta_1$) and two specific values of $k$ (i.e., $k_0$, $k_1$), as given in Table IV, it is conceivable that adequate fits may be obtained using a single gamma distribution involving only three parameters ($N$, $\theta$, and $k$), where $k$ is not necessarily integer or half-integer value. Comparing fits with a single gamma distribution vs. a sum of two gamma distributions may be made possible through use of recent availability of codes (e.g., [13]) to test nuclear models with complex calculations with globally determined parameter sets. It is also noted that $R$-matrix description of three-body final states from tritons on tritium simplifies considerably when the target is a heavy nucleus [14]. In the present case of neutrons from tritons on gold, it may be instructive to consider how this formalism may be adaptable to interpretation in terms of three-body final states. In conclusion, the present case will provide a means of generating cross sections where there are no data and where there are no experimental data to test model predictions.

**Acknowledgement:** The encouragement by members of the Nuclear Data Section of IAEA, Vienna to provide these numerical data for inclusion into the EXFOR library is very much appreciated.

**Note added in proof:** It is noted that the present 0-degree differential cross sections are also adequately described at each of the five triton energies by the sum of two functions of neutron energy $E$,

$$f_M(E) + f_F(E) = 2A\sqrt{\frac{E}{\pi}}\left(\frac{1}{\theta}\right)^{3/2}\exp\left(\frac{-E}{\theta}\right) + B \cdot \frac{E \cdot \sinh(E_F/\Gamma_0)}{\cosh(E/\Gamma_0) + \cosh(E_F/\Gamma_0)}$$

where $f_M(E)$ is a Maxwellian energy distribution that represents neutron evaporation characterized by nuclear temperature $\theta$ [12], and $f_F(E)$ is a Fermi function rising linearly with $E$ and rolling off with width $\Gamma_0$ through 0.5 maximum value at $E_F$. The function $f_F(E)$ represents emission of two correlated neutrons in the forward direction with low relative momentum [14]. This description involves a total of five parameters.

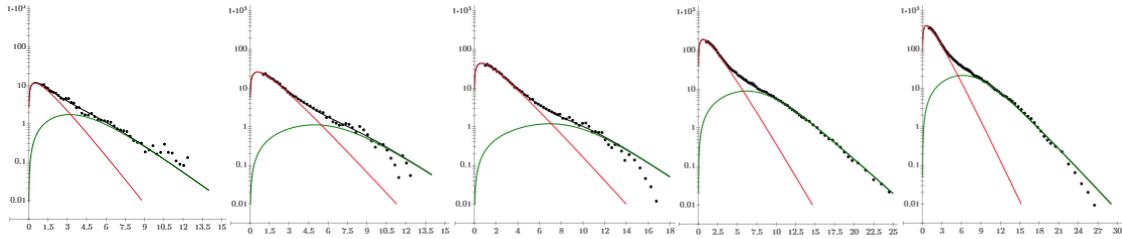

| ³H (MeV) | **5.97** | **7.47** | **10.45** | **16.41** | **19.14** |
|---|---|---|---|---|---|
| Error of fit | 0.99 | 1.1 | 2.1 | 1.9 | 4.7 |

**FIGURE IA** and **TABLE IA.** Experimental and calculated 0-degree cross sections (black points and black curves, respectively), neutron evaporation (red curves), and di-neutron emission (green curves) components, and errors of fit for five triton energies. Error of fit = 1 if deviation between measured and calculated quantities is equal to uncertainty.

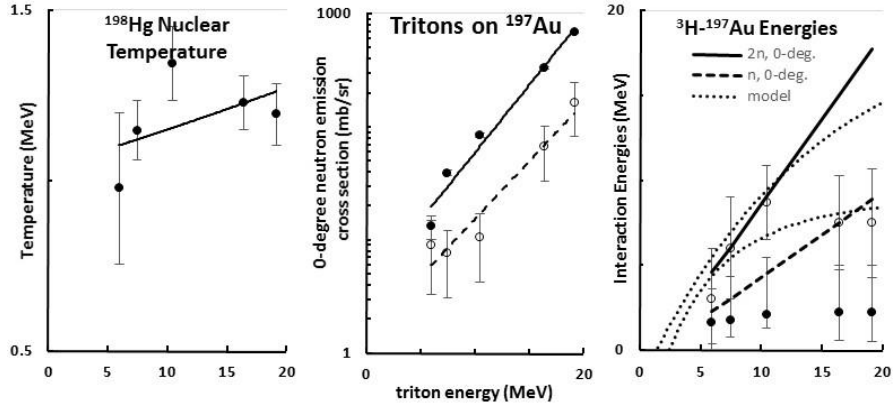

**FIGURE IIA.**

Left panel of Fig. IIA shows no significant dependence on triton energy of $^{198}$Hg nuclear temperature $\theta$ derived from 0-degree cross sections. Solid line is linear trend. Mean $\theta = 1.18 \pm 0.13$ MeV is within range of estimates of nuclear temperature in heavy nuclei.

The middle panel shows energy dependence of 0-degree neutron emission cross section for tritons on $^{197}$Au for neutrons with energy greater than 1 MeV. Cross section for evaporated neutrons is shown by solid circles. Cross section for di-neutron emission is shown by open circles. Exponential fits to the data are shown by solid lines.

The right panel shows two-body kinematics leaving $^{198}$Hg in its ground state, which results in 2n emitted at 0 degrees with maximum kinetic energy ranging, respectively, from 4.6 MeV for 5.97-MeV tritons to 17.7 MeV for 19.14-MeV tritons (solid line). Two neutrons emitted in correlation, detected separately, would therefore each have half the above kinetic energies, i.e., 2.3 MeV for 5.97-MeV tritons to 8.9 MeV for 19.14-MeV tritons (dashed line). Dependence of $E_F$ (open circles) on triton energy $E$ is shown together with a model, $E_n(E) = E_{n0}[1 - e^{(-(E-E_{th})/\Gamma_n)}]$ (dotted curves), where interaction energy $E_n(E)$ increases from zero at $^{197}$Au(t,2n) threshold energy $E_{th}$ (= 1.41 MeV) to maximum value $E_{n0}$, with width $\Gamma_n$. Two dotted curves are calculated with $E_{n0} = 20$ MeV, $\Gamma_n = 14.2$ MeV (upper curve) and $E_{n0} = 10$ MeV, $\Gamma_n = 7.1$ MeV (lower curve). Increase of $\Gamma_0$ (solid circles) with triton energy is not significant. Mean $\Gamma_0 = 1.99 \pm 0.27$ MeV.

## V. REFERENCES

[1] G. Wallerstein et al., Rev. Mod. Phys. 69, 995 (1997) http://dx.doi.org/10.1103/RevModPhys.69.995; S.J. Rose, Physics 7, 13 (2014) https://physics.aps.org/articles/pdf/10.1103/Physics.7.13; P. Navratil and S Quaglioni, Phys. Rev. Lett. 108 042503 (2011) http://dx.doi.org/10.1103/PhysRevLett.108.042503

[2] M. Drosg, G.F. Auchampaugh, and F. Gurule, Neutron Background Spectra and Signal-to Background Ratio for Neutron Production Between 10 and 14 MeV by the Reactions 3H(p,n)3He, 'H(t, n)3He, and 2H(d, n)3He, Los Alamos National Laboratory Report LA-6459-MS, August 1976.
http://permalink.lanl.gov/object/tr?what=info:lanl-repo/lareport/LA-06459-MS

[3] M. Drosg, "DROSG-2000v2.21: Neutron Source Reactions. Data files with computer codes for 59 accelerator-based two-body neutron source reactions", documented in the IAEA report IAEA-NDS-87 Rev. 9 (May 2005), IAEA, Vienna, see http://www-nds.iaea.org/drosg2000.html and http://www.nea.fr/abs/html/iaea1401.html

[4] M. Drosg: "Accurate Measurement of the Counting Efficiency of a NE-213 Neutron Detector between 2 and 26 MeV", Nucl.Instr.Meth. 105, 573 (1972) http://www.iop.org/EJ/ref/-target=doi/0305-4616/11/3/010/6

[5] M. Drosg, D.M. Drake, and P.W. Lisowski: "The Contribution of Carbon Interactions to the Neutron Counting Efficiency of Organic Scintillators", Nucl.Instr.Meth. 176, 477 (1980)
http://dx.doi.org/10.1016/0029-554X(80)90372-9

[6] PSTAR, Energy Loss Tables for Protons, as found under:
http://physics.nist.gov/PhysRefData/Star/Text/PSTAR.html

[7] MCNPX, a general-purpose Monte Carlo N-Particle eXtended radiation transport code, Los Alamos National Laboratory; https://mcnpx.lanl.gov/.

[8] Q-value calculator, http://www.nndc.bnl.gov/qcalc/qcalcr.jsp

[9] M. Drosg and D. M. Drake: "Neutron Emission Spectra of Triton Beams of 20.22 MeV Fully Stopped in Targets of $H_2O$, $D_2O$, LiF, Si, Ni, Mo, Ta, W, Pt and Au", Nucl.Sci.Eng., accepted for publication, MS# NSE15-17

[10] V.F. Weisskopf and D.H. Ewing, Phys. Rev. 57 (1940) 472-485, 935
http://dx.doi.org/10.1103/PhysRev.57.472

[11] A.Yu. Doroshenko, V.M. Piksaikin, M.Z. Tarasko: "The Energy Spectrum of Delayed Neutrons from Thermal Neutron-induced Fission of 235U and its Analytical Approximation," International Atomic Energy Agency, International Nuclear Data Committee, Vienna (Austria), Dec 2002; p. 107-113; INDC(CCP)--432;
http://www.iaea.org/inis/collection/NCLCollectionStore/_Public/34/010/34010920.pdf

[12] C. H. Holbrow and H. H. Barschall, "Neutron Evaporation Spectra." Nucl. Phys. 42, 264 (1963),
http://dx.doi.org/10.1016/0029-5582(63)90734-X

[13] A.J. Koning, S. Hilaire and M.C. Duijvestijn, ``TALYS-1.0'', Proceedings of the International Conference on Nuclear Data for Science and Technology - ND2007, April 22-27, 2007, Nice, France, eds. O. Bersillon, F. Gunsing, E. Bauge, R. Jacqmin and S. Leray, EDP Sciences, 2008, p. 211-214;
http://www.talys.eu/

[14] C.R. Brune et al., Phys. Rev. C 92, 014003 (2015), http://dx.doi.org/10.1103/PhysRevC.92.014003